\documentclass[12pt]{article}
\usepackage[breaklinks=true]{hyperref}
\usepackage{graphicx}
\usepackage{epsfig}
\usepackage{dcolumn}
\usepackage{subfigure}
\usepackage{amsmath,amsthm, amssymb}
\usepackage{comment} 
\usepackage{array}

\newcommand{\bea}{\begin{eqnarray}}
\newcommand{\eea}{\end{eqnarray}}
\newcommand{\beq}{\begin{equation}}
\newcommand{\eeq}{\end{equation}}

\def\be{\begin{equation}}
\def\ee{\end{equation}}
\def\beq{\begin{eqnarray}}
\def\eeq{\end{eqnarray}}

\newcommand{\C}[1]{\mathcal{#1}}

\begin{document}

\begin{center}
\vspace{48pt}
{ \Large \bf An Analytical Analysis of CDT \\  Coupled to Dimer-like Matter}

\vspace{40pt}
{\sl Max R. Atkin}$^{a}$
and {\sl Stefan Zohren}$\,^{b}$  
\vspace{24pt}

{\small

$^a$~Fakult\"{a}t f\"{u}r Physik, Universit\"{a}t Bielefeld,\\
Postfach 100131, D-33501 Bielefeld, Germany

\vspace{10pt}

$^b$~Rudolf Peierls Centre for Theoretical Physics,\\
1 Keble Road, Oxford OX1 3NP, UK.
}

\end{center}


\vspace{36pt}

\begin{center}
{\bf Abstract}
\end{center}
We consider a model of restricted dimers coupled to two-dimensional causal dynamical triangulations (CDT), where the dimer configurations are restricted in the sense that they do not include dimers in regions of high curvature. It is shown how the model can be solved analytically using bijections with decorated trees. At a negative critical value for the dimer fugacity the model undergoes a phase transition at which the critical exponent associated to the geometry changes. This represents the first account of an analytical study of a matter model with two-dimensional interactions coupled to CDT. 


\vfill

{\footnotesize
\noindent
$^a$~{email: matkin@physik.uni-bielefeld.de}\\
$^b$~{email: zohren@physics.ox.ac.uk}\\

}


\newpage

\section{Introduction}
The program of Causal Dynamical Triangulations (CDT) is an attempt to make the notion of quantum gravity as the path integral over geometries well defined \cite{Ambjorn:1998xu}. In this approach one approximates the path integral by a sum over discrete approximations to each geometry known as causal triangulations. Causal triangulations have a preferred time foliation with respect to which spatial topology change is forbidden.

An unfortunate aspect of the casual restriction of CDT, compared to the earlier Dynamical Triangulations (DT) approach \cite{Ambjorn:1997di}, is that the relation to matrix models, as well as to a continuum path integral formulation is lost - although there has been progress in establishing a matrix model formulation \cite{Ambjorn:2008ta,Ambjorn:2008gk,Dario}
as well as relations to a continuum path integral in the proper time gauge \cite{Westra:2011rg} (see also \cite{Ambjorn:2009rv} for a review). Due to the loss of the matrix model formulation, the original study of CDT used a transfer matrix method to solve the pure case in two dimensions and in higher dimensions all work has been numerical in nature (see \cite{Ambjorn:2011kp} for a recent review). This has the consequence that unlike DT, it has been very difficult to analytically investigate matter coupled to CDT. One exception to this was the work of \cite{Di Francesco:2001gm}, where a peculiar form of dimer matter was studied in which dimers were allowed to be of arbitrary length in the timelike direction. Although they were able to obtain a multicritical point, this model was in some sense one-dimensional, due to the absence of spacelike interactions between dimers and it is unclear what sort of theory the continuum limit represents. Despite this absence of analytical work, there have been a number of very interesting numerical studies of the behaviour of various forms of matter coupled to two-dimensional CDT \cite{Ambjorn:1999gi}.

These studies indicated two intriguing properties; firstly, unlike DT, CDT does not appear to modify the critical exponents of the CFT and secondly, the $c=1$ barrier appears to be of a milder form with indications the geometry instead changes to a ``blob'' type geometry of higher dimension \cite{Ambjorn:1999yv}.
These properties are currently conjectures based on numerical evidence and therefore it is a pressing open problem in the study of CDT to find a method by which matter coupled to CDT can be analysed analytically.

In this paper we introduce a method of analytically computing the partition function for a certain kind of hard dimer model coupled to CDT. Our method is in the spirit of the tree bijection techniques used by \cite{Schaeffer-Ising}
in which it was shown that conformal matter coupled to DT could be analysed using such bijections without ever referring to matrix integrals.



\section{Hard Dimers in DT and CDT}
The particular matter model we will consider is known as hard dimers. For a regular triangular lattice a hard dimer model consists of objects known as dimers which occupy two adjacent sites in the lattice. Each dimer present in the lattice contributes a factor of 
 $\xi$ to the weight associated to the entire configuration. The interaction between the dimers is via a ``hard'' interaction in which only a single dimer can reside inside a given triangle. This property makes dimers somewhat easier to analyse than other models such as the Ising model, as the interaction only manifests itself through the number of valid configurations. 

\subsection{Dimers in DT}
We first review how to couple dimers to DT by making use of matrix model techniques \cite{Kazakov:1989bc,Staudacher:1989fy}. Since matrix models generate the dual graph of the triangulations it is easiest to proceed by reformulating the dimer model on the dual graph. Such a reformulation is simple; the model corresponds to colouring the links of the graph with two colours, black and red, such that red links can never end on the same vertex. The red links then correspond to the dual of the dimers.

A model which generates Feynman diagrams appropriate for counting such a system is given by the Hermitian two-matrix model for matrices of size $N\times N$,
\beq
\label{0mm}
Z = \int [dM] [dA] e^{-N \mathrm{tr}[ \frac{1}{2} M^2 - \frac{g}{3} M^3 + \frac{1}{2} A^2 - g \sqrt{\xi} M^2 A ]}
\eeq
where $A$ generates red links and $M$ black. In this model $g$ is the fugacity associated to a single triangle and $\xi$ is the weight associated to the inclusion of a dimer. In the continuum limit $g$ will scale to a coupling describing a perturbation by the lowest dimension operator present in the continuum theory. Usually such an operator is the identity operator and therefore $g$ may be interpreted as the cosmological constant $\Lambda$. However, for non-unitary theories there exist negative dimension operators and $g$ can no longer be given this interpretation. The model contains a critical point $(g_c, \xi_c)$ about which a scaling limit can be defined which is described by the $(2,5)$ minimal CFT coupled to gravity which is non-unitary. The observables of this model are $\gamma$, known as the string susceptibility, and defined as the first non-analytic power in,
\beq
\label{gammadef}
Z \sim \Lambda^{2-\gamma},
\eeq
and the critical exponent $\sigma$, defined similarly through \cite{Staudacher:1989fy},
\beq\label{defsigma}
\frac{d \log{g_c(\xi)}}{d \xi} \sim (\xi - \xi_c)^\sigma.
\eeq
Note that for models in which the identity operator is the lowest dimension operator, \eqref{gammadef} can be written as,
\beq
\label{gammadef2}
Z \sim (g-g_c)^{2-\gamma}.
\eeq
For models in which the identity is not the lowest dimension operator we call the exponent associated with \eqref{gammadef2} the modified string susceptibility. The string susceptibility characterises the modification of the geometry by the presence of matter. By solving the model \eqref{0mm} it was found that \cite{Staudacher:1989fy},
\beq
\gamma = -\frac{3}{2}, \qquad \gamma_\mathrm{modified} = -\frac{1}{3}, \qquad \sigma = \frac{1}{2}.
\eeq
This should be contrasted with the case of dimers on a flat regular lattice, for which $\sigma = -\frac{1}{6}$.

\begin{figure}[t]
\centering 
\includegraphics[scale=0.4]{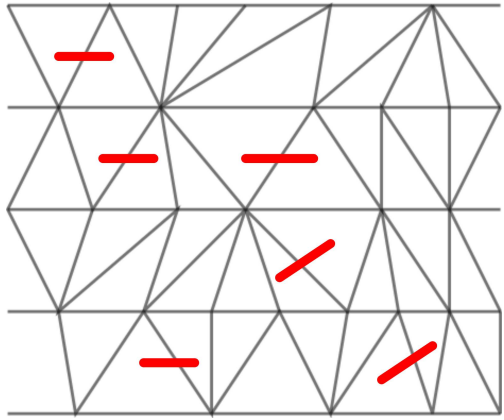}
\caption{A configuration of a CDT coupled to dimers as considered in \cite{DiFrancesco:1999em}: There are no interactions between adjacent spatial slices.}
\label{CDTwithdimers}
\end{figure}

\subsection{Dimers in CDT}
There has been one previous attempt to study CDT coupled to the same sort of hard dimer matter as found in DT \cite{DiFrancesco:1999em}. However, the result of this study was that although the discrete model could be solved, the scaling limit was still in the universality class of pure CDT. Before moving on to our own construction it is instructive to understand the failure of this previous study.
In \cite{DiFrancesco:1999em} a generalisation of CDT was analysed in which a discrete version of a higher curvature term was added to the measure. This corresponded to attaching an extra weight to each instant in a triangulation when two adjacent triangles in the same space-like slice are of the same orientation. It was shown in \cite{DiFrancesco:1999em} that this weighting maps to a dimer-like system in which the dimers are confined to lie only across time-like edges connecting an up- and a down-pointing triangle, resulting in configurations similar to that shown in Figure \ref{CDTwithdimers}.

For statistical models such as the Ising or dimer model, the critical point, at which the model is described by a conformal field theory, corresponds to the situation in which the correlation length diverges. It is therefore perhaps not surprising that the dimer-like model considered in \cite{DiFrancesco:1999em} was still in the universality class of pure gravity; there is no correlation of dimer configurations even between adjacent layers. This effectively transform the problem into a succession of independent dimer models on one-dimensional circles, for which it is known that no phase transition exists. It is clear then that in order to move out of the universality class of pure CDT we must at least allow for some interaction between adjacent layers. This is the subject of the next section.

\section{Decorated Trees and Dimers}
It has been known for some time beginning with the work of Cori-Vauquelin \cite{Cori} and Schaeffer \cite{Sch:1998} that there exist bijections between tree graphs and various kinds of triangulations (not necessarily composed of triangles), known in the mathematical literature as maps.
A particularly interesting use of these bijections have been the solutions of the Ising model and other models \cite{Schaeffer-Ising}
coupled to two-dimensional gravity without the use of matrix models. The purpose of this paper is to apply these techniques to the problem of hard dimers coupled to CDT.

\subsection{A Tree Bijection for Pure CDT}
\begin{figure}[t]
\centering 
\parbox{7cm}{
\includegraphics[scale=0.5]{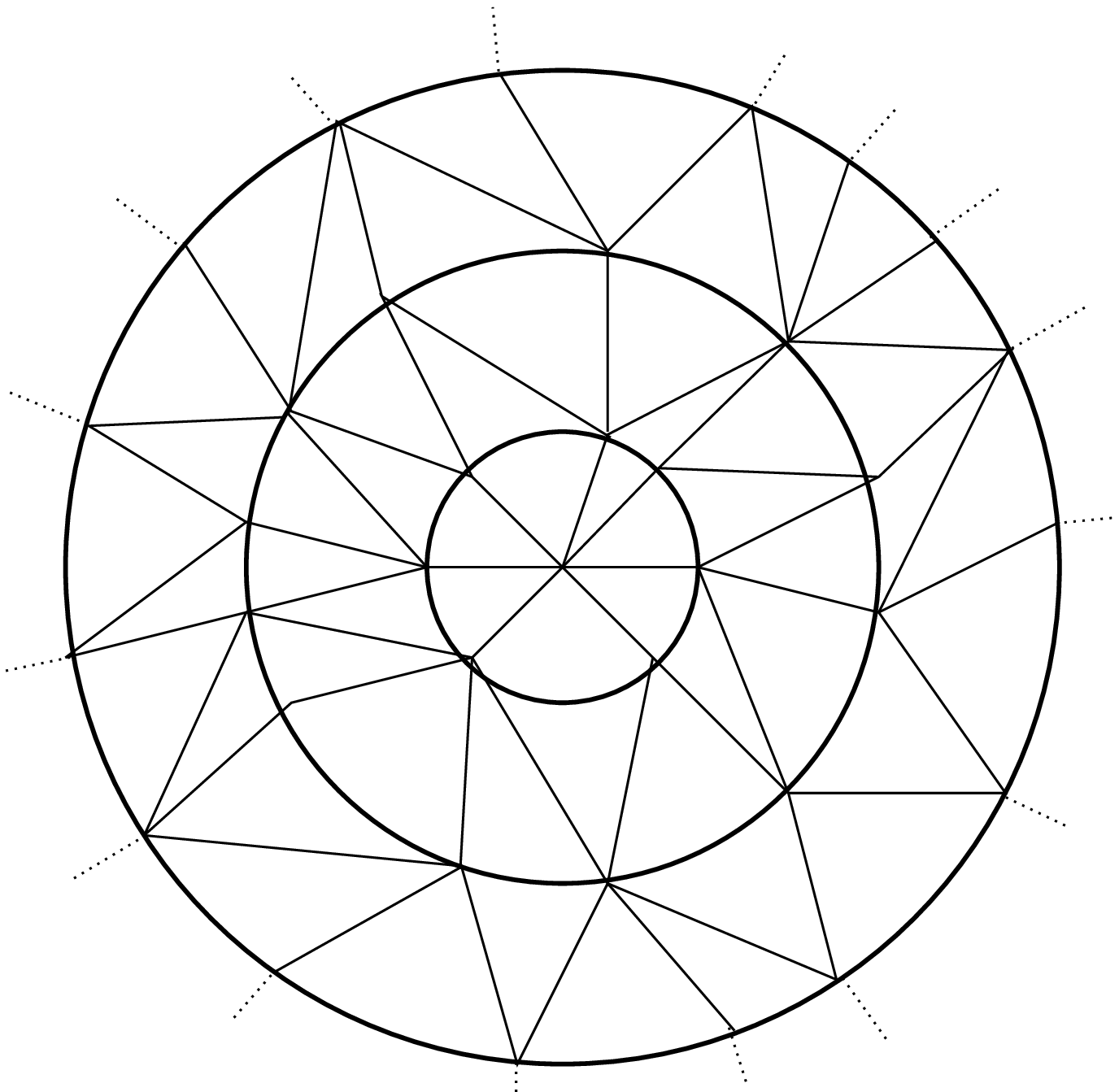}
\caption{A causal triangulation drawn in the plane. The dashed links emanating from the outermost circle all end on a single vertex in the infinite face, which we have omitted from the diagram.}
\label{CDTbijection1}}
\qquad 
\begin{minipage}{7cm}
\includegraphics[scale=0.5]{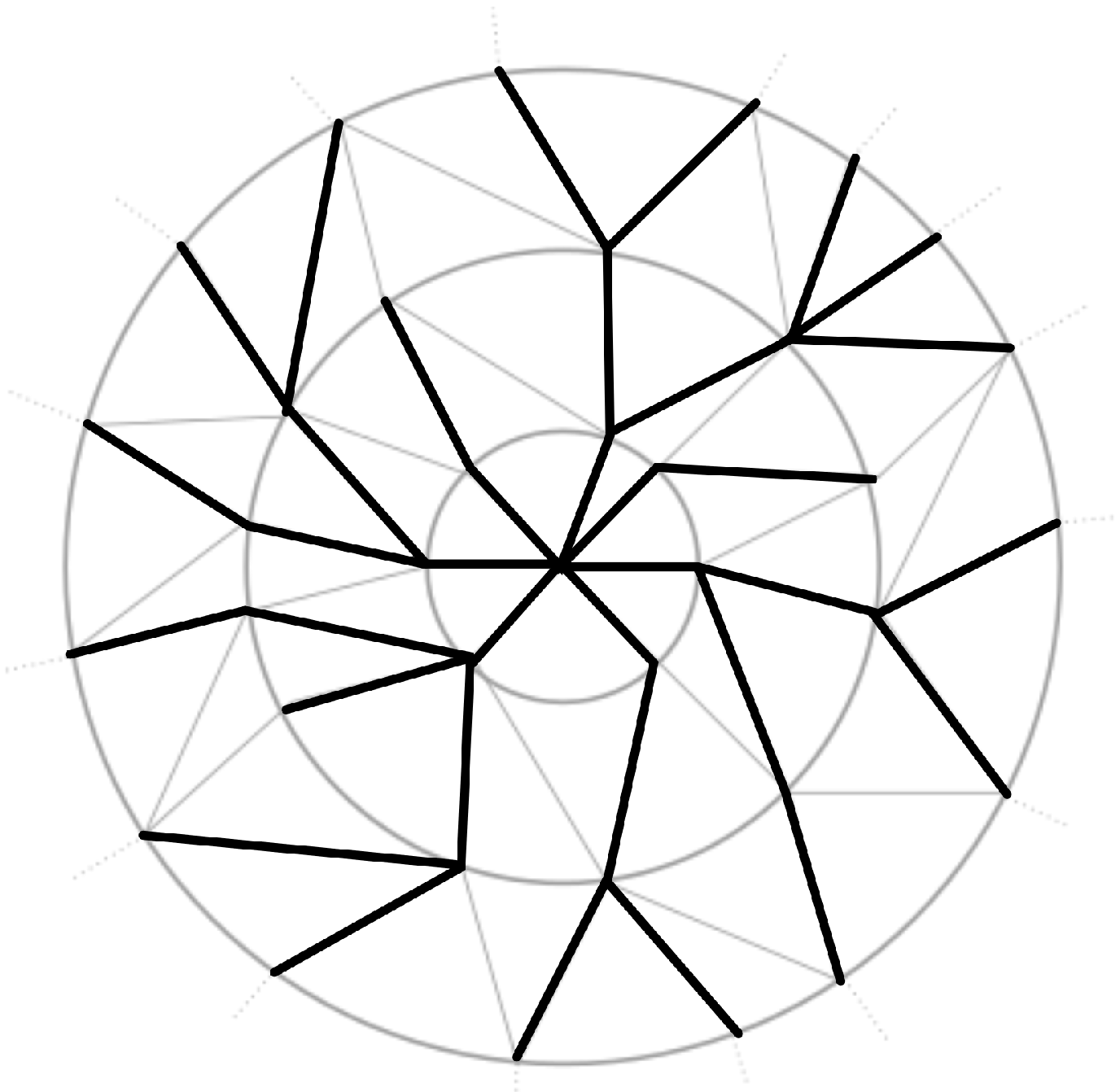}
\caption{Causal triangulations have a bijection with trees. In this case the tree associated to the triangulation is shown by thick lines. We have also omitted the root vertex attached to the central vertex.}
\label{CDTbijection2} 
\end{minipage} 
\end{figure}

The bijections formulated in \cite{Sch:1998,bouttier-2002-645} related maps to various kinds of labelled trees. The maps themselves were precisely of the kind that contribute to the partition function of DT. To apply these techniques we first review a tree bijection from CDT to rooted planar\footnote{Here ``planar'' refers to the fact that the embedding of the tree in the plane is significant, in that it determines an order on the vertices in that one proceeds another in an (anti) clock-wise direction in the plane.} trees. Such a bijection was first introduced in \cite{DiFrancesco:1999em} in a form differing from the one used here. The bijection used here was introduced by \cite{Durhuus:2009sm,MYZ2001} and has been used for example to bound the spectral dimension of the spacetimes resulting from CDT \cite{Durhuus:2009sm}. The bijection is most easily described by first flattening the CDT graph for a cylinder on to the plane, as is shown in Figure \ref{CDTbijection1}. Next, we denote the vertices in the $k$th spacelike slice as $S_k$, we define $S_0$ as containing a single new vertex connected to all vertices in $S_1$. Following the formulation in \cite{Durhuus:2009sm}, the following steps then map the CDT to a rooted planar tree $T$;

\begin{itemize}
\item[1.] All vertices in the triangulation are in $T$ in addition to a new root vertex, called the root, that is only linked to $S_0$.
\item[2.] All links from $S_0$ to $S_1$ are in $T$.
\item[3.] All links from a vertex in $S_k$ to vertices in $S_{k+1}$, apart from the anticlockwise-most link, are in $T$.
\end{itemize}
These steps produce the tree shown in Figure \ref{CDTbijection2}. Denoting the vertices at a height $n$ above the root by $V_n$, this procedure can also be shown to be invertible by the following steps,
\begin{itemize}
\item[1.] Delete the root node and the link joining it to $V_1$. We identify $V_1$ as $S_0$. 
\item[2.] For all $n < h$, where $h$ is the total height of the tree, join each vertex in $V_{n+1}$ to the next vertex in the clockwise direction, as determined by the embedding in the plane.
\item[3.] For each vertex $v$ in $V_n$, $1<n<h$, that is not of order one in the tree, add a link between $v$ and a vertex in $V_{n+1}$ such that the new link is the most anticlockwise link emerging from $v$ and furthermore maintains the planarity of the graph.
\item[4.] For vertices of order one, add a new link between it and the unique vertex in the next layer such that the planarity of the graph is maintained.
\end{itemize}

\subsection{A Tree Bijection for a Restricted Dimer Model}
\begin{figure}[t]
\centering 
\parbox{7cm}{
\includegraphics[scale=0.5]{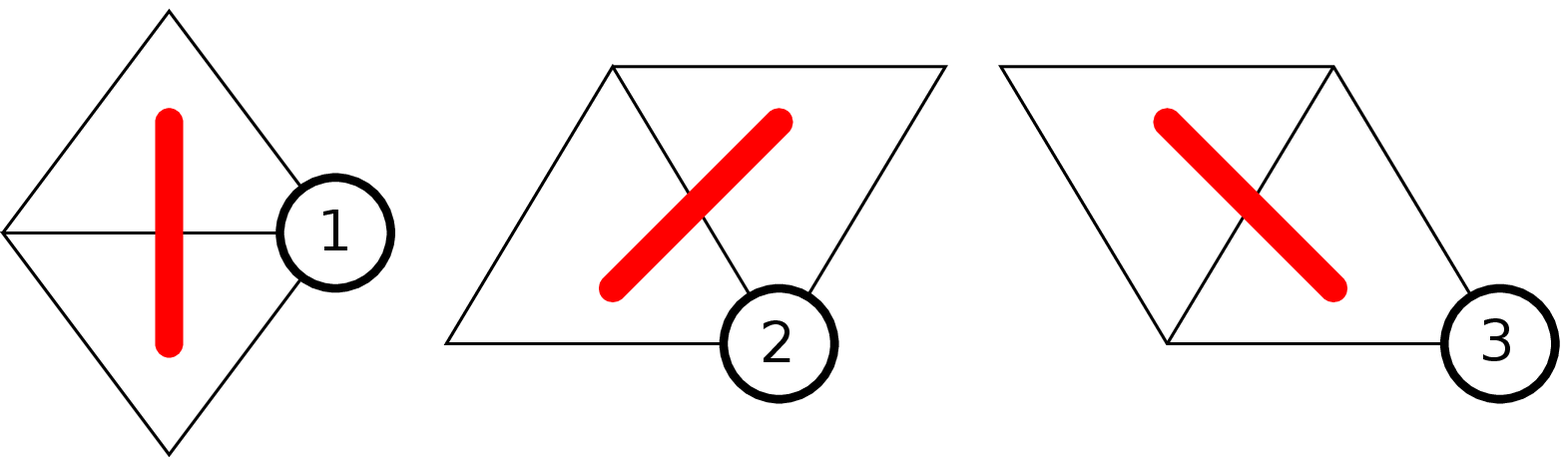}
\caption{The dimers configurations that are permitted in the restricted dimer model.}
\label{dimerconfig}}
\qquad 
\begin{minipage}{7cm}
\includegraphics[scale=0.5]{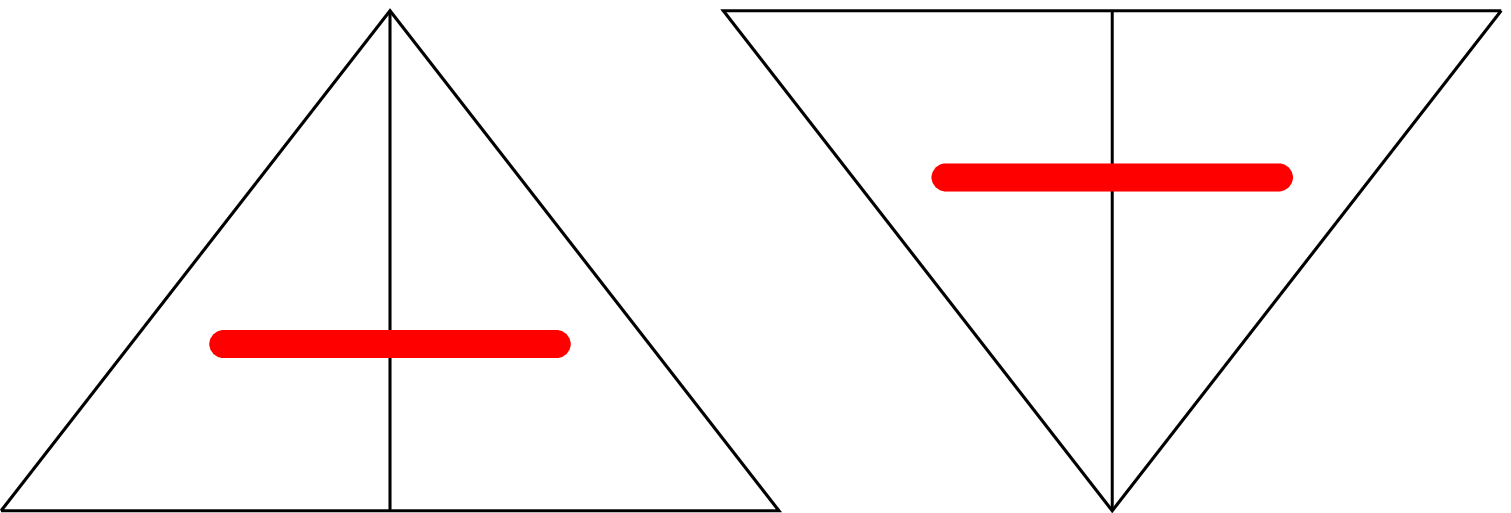}
\caption{The dimer configurations that are excluded in the restricted dimer model.}
\label{dimerconfig2} 
\end{minipage} 
\end{figure}

We now present an extension of the above bijection to allow certain dimer configurations to be encoded in the tree. Given a triangulation, we first label all vertices with the number zero. Then consider Figure \ref{dimerconfig}; it displays a number of possible ways a dimer could be embedded into a triangulation. Note that the list of how a dimer may be embedded in the triangulation is not exhaustive and neglects the two possibilities shown in Figure \ref{dimerconfig2}. We refer to the model which neglects these possibilities as the restricted dimer model. Note that our model includes the dimer configurations of \cite{DiFrancesco:1999em} as a subset in addition to allowing dimers to lie between layers, as is required to move out of the universality class of CDT.

\begin{figure}[t]
\centering 
\parbox{7cm}{
\includegraphics[scale=0.5]{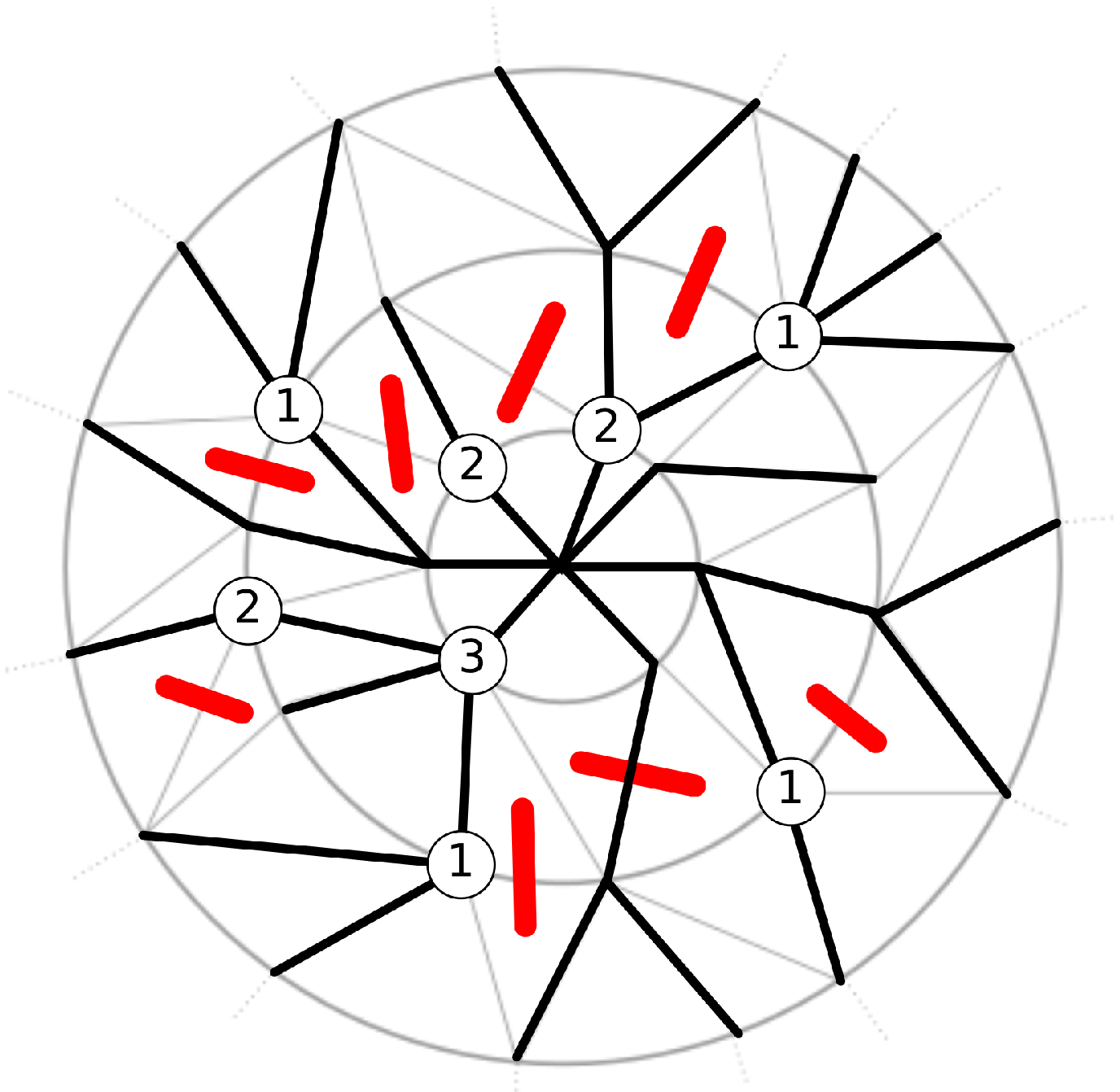}
\caption{An example of the labelled tree encoding the dimer configuration in the CDT.}
\label{labelledtree}}
\qquad 
\begin{minipage}{7cm}
\includegraphics[scale=0.5]{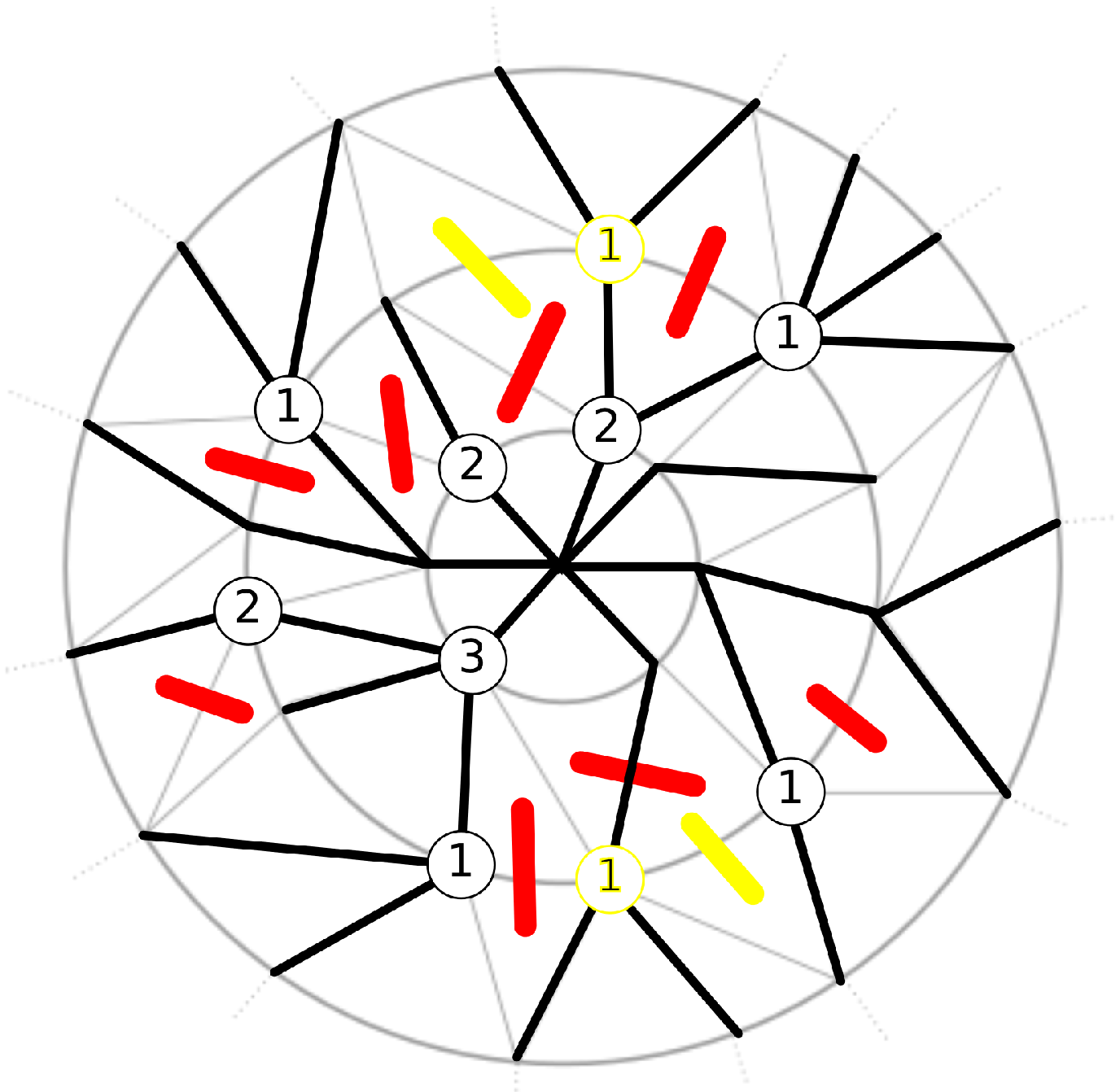}
\caption{Here, in yellow, we see two invalid labels in the tree. At the top of the diagram we see why a type $2$ label may not have a type $1$ label as its anti-clockwise most child. At the bottom of the diagram we see how the label $3$ causes there to be non-local constraints on the labelling.}
\label{badtree} 
\end{minipage} 
\end{figure}

When one of the restricted dimer configurations appears in a triangulation we will label the vertex shown in Figure \ref{dimerconfig} with a number corresponding to how the dimer is configured. By noting that the tree bijection of the previous section does not delete any vertices, we can extend it to the present case simply by defining it to preserve the labelling of the vertices, thereby obtaining a labelled tree, Figure \ref{labelledtree}.

We now address the question of what class of labelled trees is produced by the above map. Firstly, not every labelled tree is in the range of the map. This fact can be seen by noting that certain labelled trees can not arise as they would correspond to a configuration of dimers in which some occupied the same site, as shown in Figure \ref{badtree} by the yellow dimers. We therefore have a tree for which certain restrictions are present on the labelling. The task is now to characterise these restrictions and furthermore find a way to count the number of trees appearing in this class. \emph{Note that in doing so we have to account for the non-local interactions between different branches of the tree.}

We begin by considering the labelling on a tree. We see that a vertex labelled with a $0$, $1$ or $3$, places no restrictions on the labelling of any of its children. However, the label $3$ will affect the next branch in an anti-clockwise direction as can be seen with the lower yellow dimer in Figure \ref{badtree}. This effect is non-local in the tree graph and we postpone a discussion on how to handle it until later. For a vertex carrying the label $2$, we see from the upper yellow dimer in Figure \ref{badtree} that it must have at least one child and furthermore this child can only be labelled with a $0$ or $2$. Furthermore such a dimer does not place any other constraints on the labelling of the graph and hence its effect is local in the tree graph. Since the rules associated with labels $0$, $1$ and $2$ only put restrictions on the child nodes of a vertex in the tree, trees containing only type $0$, $1$ and $2$ dimers can be counted recursively via generating functions.

The type-$3$ dimers may be included by mapping them onto type-$2$ dimers via another bijection. The type-$2$ and type-$3$ dimers have the special property that by flipping the central diagonal we may convert one in to another, as in Figure \ref{23flip}, and crucially, {\emph{this does not affect any other dimers in the configuration}}. To clarify, by performing the move in Figure \ref{23flip} we map one triangulation with a particular configuration of dimers to a distinct triangulation with a distinct configuration of dimers containing one less type-$2$ dimer and one more type-$3$ dimer. Hence, {\emph{by counting each type-$2$ dimer twice we are able to account for all configurations including type-$3$ dimers}. This is a crucial non-trivial observation necessary to perform the calculation in this paper.\footnote{For the sceptical reader, we have verified that our generating function derived via this argument reproduces the correct number of restricted dimer configurations for triangulations of up to size 18 as found by computer.}

\begin{figure}[t]
\centering 
\parbox{7cm}{
\includegraphics[scale=0.4]{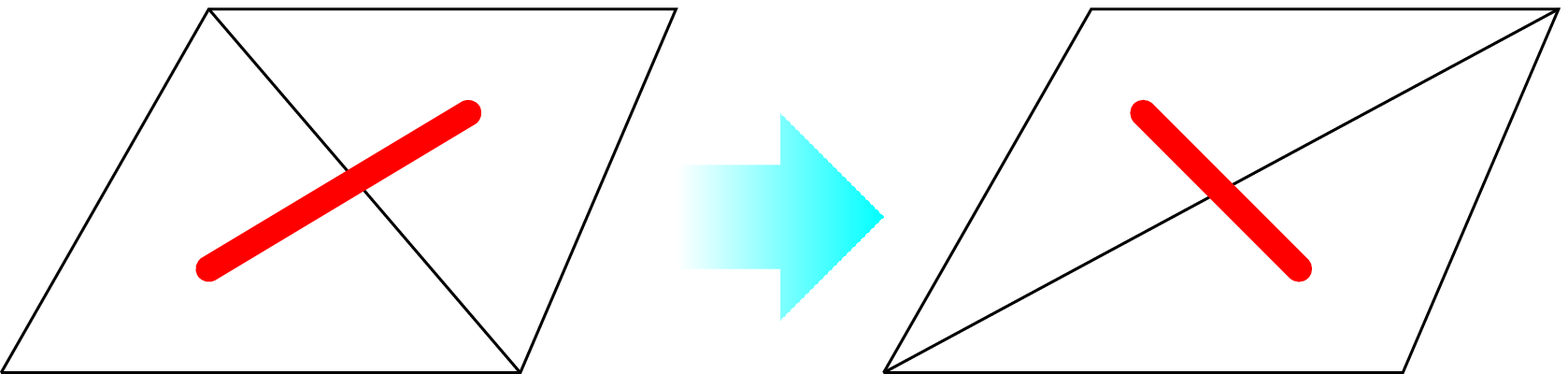}
\caption{The move used to transform a type $2$ dimer into a type $3$ by changing the triangulation. This move can be performed independently of the rest of the triangulation.}
\label{23flip}}
\qquad 
\begin{minipage}{7cm}
\includegraphics[scale=0.4]{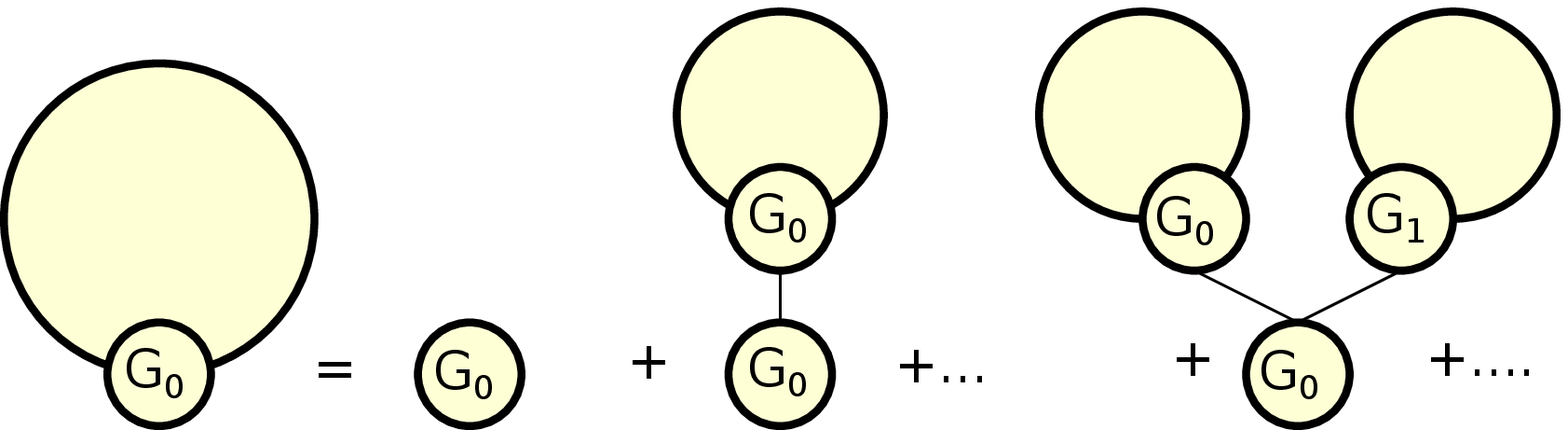}
\caption{A graphical representation of the recursion relation needed to compute the generating function for the labelled trees.}
\label{G0eqn} 
\end{minipage} 
\end{figure}

We are now in a position to count the number of labelled trees satisfying the above rules. Let us introduce the generating functions $G_i$ defined by,
\beq
G_i(\xi,g) = \sum_{m=0}^\infty\sum_{n=0}^\infty \C{N}_i(n,m) g^{2n} \xi^m,
\eeq
where $\C{N}_i(n,m)$ is the number of trees with a root carrying the label $i$, containing $n$ vertices, $m$ of which carry a non type-$0$ label, thereby indicating the presence of $m$ dimers. We can obtain an expression for $G_0$ of the form,
\beq
G_0(\xi,g) = \frac{1}{1-g^2(G_0 + \xi G_1 + 2\xi G_2 )} = 1 + g^2(G_0 + \xi G_1 + 2\xi G_2 ) + \ldots,
\eeq
which is shown pictorially in Figure \ref{G0eqn}. Given that the type-$1$ label places no restriction on the children we also have $G_1 = G_0$. Lastly, by recalling the restrictions due to a type-$2$ dimer, we can write,
\beq
G_2(\xi, g) = g^2(G_0 + 2\xi G_2)G_0.
\eeq
Solving for $G_0$ results in the equation,
\beq
\label{Geqn}
2 g^4  \xi^2 G_0^3 - g^2 (1 + 3 \xi)G_0^2 + (1 + 2 g^2 \xi)G_0 -1 =0.
\eeq
The critical line $g_{cr}(\xi)$ of this model, at which the mean size of the triangulations diverges, can be found from the radius of convergence of $G_0$. This is easily found by computing the discriminant of \eqref{Geqn}, obtaining the following equation for $g_{cr}(\xi)$,
\beq
\label{gcreqn}
64 \xi^5 g_{cr}^6 - 4 \xi^2 (1 + 24 \xi + 12 \xi^2)g_{cr}^4 +  4 (1 + 8 \xi + 12 \xi^2 + 3 \xi^3)g_{cr}^2 -1 - 6 \xi - \xi^2= 0.
\eeq
From the discriminant of the above equation there exist two possible critical values of $\xi$; $\xi_{cr} = \pm (27)^{-1/2}$. Furthermore we identify the correct branch of $g_{cr}$ by requiring that it reproduces for $\xi=0$ the known critical value of $g_{cr}(0) = 1/2$ for pure CDT.

By considering how the roots of \eqref{gcreqn} change as $\xi$ varies, one can see that there exists one branch that approaches $g_{cr}= 1/2$ as $\xi \rightarrow 0^+$ and another which approaches $g_{cr} = 1/2$ as $\xi \rightarrow 0^-$, with each solution diverging and becoming complex upon passing through $\xi = 0$. If we identify these two branches at $\xi = 0$ then we obtain the plot of $g_{cr}$ show in Figure \ref{gcrplot}.
 
\begin{figure}[t]
\centering 
\includegraphics[scale=0.9]{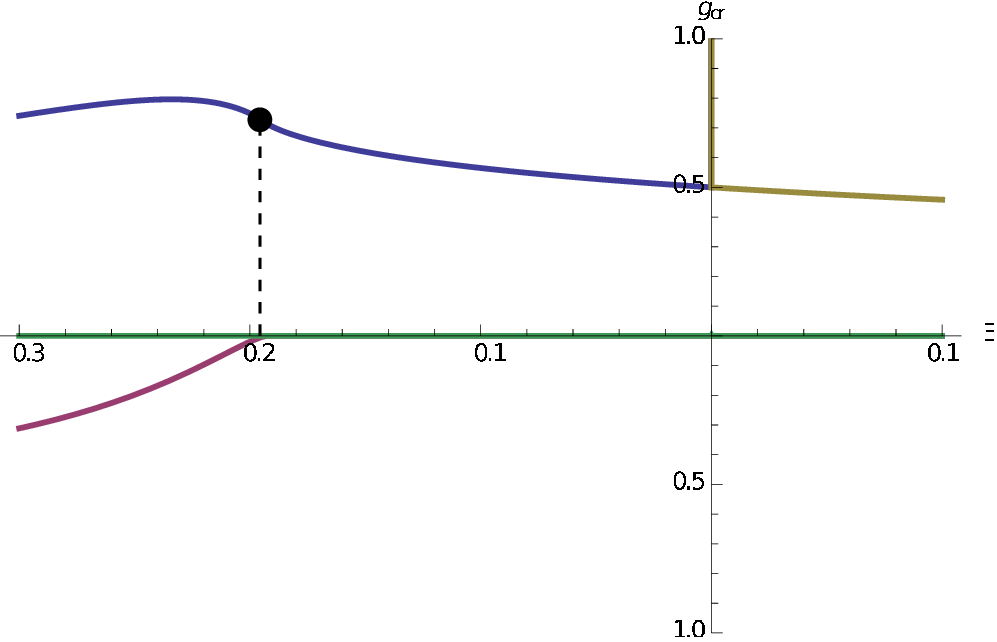}
\caption{A plot of the value of $g_{cr}$. Its real value is shown in blue and gold, corresponding to the two branches identified at $\xi = 0$. The imaginary value of the respective branches is shown in red and green. One can see that the critical line ends at the value of $\xi_{cr} = -(27)^{-\frac{1}{2}}$, marked by the dot. }
\label{gcrplot}
\end{figure}
Expanding the solution in a taylor series around its endpoint at $\xi_{cr}$ we obtain using \eqref{defsigma},
\beq
\sigma = \frac{1}{2},
\eeq
and furthermore substituting the critical value $\xi$ into the solution for $G_0$ we obtain,
\beq
\gamma_{\mathrm{modified}} = \frac{1}{3},
\eeq
from which it is clear that we are now outside the universality class of pure CDT! 
These results are quite striking when compared to the flat space and DT exponents. Firstly, the value of $\gamma_{\mathrm{modified}}$ is positive. For all minimal models coupled to DT $\gamma$ is negative and historically a positive $\gamma$ was associated with going beyond the $c = 1$ barrier \cite{Ambjorn:1994ab}.  

Secondly, $\sigma$ has exactly the same value as obtained in DT. This is particularly surprising given the numerical evidence that coupling to CDT preserves the flat space exponents. One should note however that all numerical work on matter coupled to CDT has used matter for which $c > 0$. Of course, one could argue that we have not solved the full dimer model and that we might obtain agreement with the numerical work once we include all configurations. However, we will now argue that even in the unrestricted case we will have $\sigma$ differing from its flat space exponent.

Note that the dimers shown in Figure \ref{dimerconfig2} correspond to dimers placed in regions of high curvature. One consequence of this is that highly curved spacetimes will be entropically suppressed. Furthermore, the negative fugacity of the dimers at their critical point will also enhance the weight associated to spacetimes that can support dimers. Hence there is a double enhancement of the flatter spacetimes. Hence, if our results for the critical exponents were to change for the unrestricted dimer model, we would expect them to move further from their flat space counter parts.
Therefore, we expect that the full dimer model coupled to CDT belongs to the same universality class as the restricted dimer model coupled to CDT as considered here.


\section{Discussion}
In this paper we have solved analytically a hard dimer model with two dimensional interactions coupled to CDT. We have been able to compute the disc function, the string susceptibility and the critical exponent for the hard dimer model, $\sigma$. Both exponents were seen to differ from the pure CDT case, hence indicating we have been able to analytically study a known matter model coupled to CDT for the first time. Furthermore, the value of $\sigma$ we found matched that obtained from DT. This is particularly surprising given the numerical evidence that matter coupled to CDT does not have critical exponents that differ from their flat space values.

The particular hard dimer model we considered is an extension of the one considered in \cite{DiFrancesco:1999em} with the crucial difference that we are able to include dimers laying both within a single spacelike slice and those which lie across two. The ability to include both types of dimers is the fundamental reason that this model lies in a different universality class from pure gravity; dimers can interfere with one another across layers, allowing for long range correlations to arise.

We build on the work of \cite{Schaeffer-Ising} 
in which bijections were constructed between labelled tree graphs and DT with matter, in addition to \cite{DiFrancesco:1999em,Durhuus:2009sm,MYZ2001} which introduced a bijection between unlabelled trees and pure CDT. In our case we demonstrated that in the case of CDT we could keep track of the dimers by introducing a labelling on the tree. This in itself is not surprising, however since the connectivity of the tree graph is lower than the triangulation, interactions between neighbouring vertices in the triangulation generically become non-local interactions in the tree. A novel aspect of our work is that we were able to account for the non-local interactions appearing in the restricted dimer model by introducing a mapping between trees. This map relates trees with only local interactions to trees with non-local ones, allowing us to count the latter.

The dimer configurations that we have not included in this model correspond to dimers embedded in regions of high curvature, thereby suppressing spacetimes which are highly curved. We therefore physically expect our model to scale to a continuum model in which there exists a coupling between matter and curvature. Furthermore, we have argued that this implies that the value of $\sigma$ will not return to its flat space value upon the inclusion of all dimer configurations.
  
\subsection*{Note} While completing this article we were informed by Jan Ambj{\o}rn about an unpublished work of him, L. Glaser, A. G{\"o}rlich and Y. Sato regarding the multi-critical analogue of a new continuum limit of matrix models as introduced in \cite{Ambjorn:2008gk} in which they obtain the same critical exponents as those in this article. This is very interesting, as the approach of their paper is in some sense highly complementary to ours. Indeed their work provides very strong evidence that the restricted dimer model belongs to the same universality class as the full dimer model, as we argued in this paper. An obvious advantage of the matrix model approach is that it is much easier to generalise to other multicritical points, however even in the case of DT it was difficult to be certain what continuum theory a particular multicritical point scaled to. The identification of a particular scaling limit with a CFT coupled to gravity was performed by matching the exponents with those from the KPZ formula. However in CDT no such formula exists and therefore our work complements the work of Ambj{\o}rn et al.\ nicely in that it gives an explicit demonstration that the multicritical point found is indeed the continuum theory associated with the hard dimer phase transition.

\subsection*{Acknowledgements} The authors would like to thank J. Ambj{\o}rn for discussions and for sharing his results with us prior to publication. MA acknowledges the financial support of Universit\"{a}t Bielefeld. SZ acknowledges financial support of the STFC under grant ST/G000492/1. Furthermore, he would like to thank the Mathematical Physics Group at Universit\"{a}t Bielefeld for kind hospitality and financial support for a visit during which this work was completed.
 


\providecommand{\href}[2]{#2}\begingroup\raggedright\endgroup

\end{document}